\documentclass[pre,onecolumn]{revtex4}
\usepackage{epsfig}
\normalsize
\makeatletter

\makeatother

\begin{document}

\title{MEMORY EFFECTS IN NONLINEAR TRANSPORT: KINETIC EQUATIONS AND RATCHET
DEVICES}

\author{I. Santamar\'{\i}a-Holek and J. M. Rub\'{\i}}

\affiliation{Departament de F\'{\i}sica Fonamental-CER F\'{\i}sica de Sistemes Complexos,
Facultat de F\'{\i}sica Universitat de Barcelona Diagonal 647, 08028 Barcelona,
Spain}

\begin{abstract}
We present a new method to derive kinetic equations for systems undergoing 
nonlinear transport in the presence of memory effects. In the framework of mesoscopic nonequilibrium
thermodynamics, we obtain a generalized Fokker-Planck equation incorporating memory effects
through time-dependent coefficients. As applications, we first discuss the non-Markovian dynamics of 
anomalous diffusion in a potential,
analyzing the validity of the fluctuation-dissipation theorem. In a second 
application, we propose a new ratchet mechanism in which the 
periodic driving acting on the particle is induced by the Onsager coupling of the 
diffusion current with an oscillating thermodynamic force.

\end{abstract}

\maketitle

\section{Introduction}

The characterization of systems outside equilibrium requires an \emph{a priori}
identification of the relevant time and length scale playing a major role in
the dynamics. Once this task has been accomplished one then proceeds to establish
the set of kinetic equations governing the behaviour of the probability densities
and subsequently to obtain the correlation functions which can be contrasted
with experiments. These are the main steps nonequilibrium statistical
mechanical theories set up to analyze the dynamics of systems of very different
nature evolving under out-of-equilibrium conditions \cite{zwanzig libro}.

The present-day-tendency in the study of systems working at micro- and nano-scales,
based on the increasing experimental possibilities, demands
the implementation of simple theoretical frameworks able to cope with the complexity
inherent to the analysis of systems with many competing time and length scales
(usually referred to as complex systems).
Two are the main characteristics exhibited by those systems. The unavoidable
presence of energetic and entropic barriers 
makes the dynamics intrinsically nonlinear \cite{ibuki}-\cite{rev-reimann}. 
The drastic elimination
of degrees of freedom usually performed for the sake of simplification is accompanied
by the appearance of memory effects \cite{victor}-\cite{cellular transport}. These two characteristics, together with
the fact that frequently the system evolves under the influence of external
gradients or fields creating an inhomogeneous environment, must be reflected
in the formulation of the kinetic equations. To implement that task, many efforts
have been devoted in the framework of kinetic theory \cite{kac-logan}, projector
operator techniques \cite{mori}-\cite{pop-mori} and in the theory of stochastic
processes \cite{oxtoby 1}-\cite{kampen}.

Two types of Fokker-Planck equations incorporating memory effects are frequently presented. The first
one introduces the non-Markovianity in terms of retarded kernels, i.e., by incorporating time
integrals in the equation \cite{zwanzig}, \cite{sokolov}. In the second type, non-Markovianity is taken
into account through a time dependence of the coefficients \cite{kac-logan}, \cite{pop-mori}, \cite{adelman},
\cite{cukier}. A detailed discussion explaining their differences together with experimental
implications can be found in \cite{marcus}.

Our purpose in this paper is to propose a new method to derive kinetic
equations of the Fokker-Planck type in which memory effects are incorporated through
time-dependent coefficients. In recent works \cite{gradTempNpart.}-\cite{beyond}, we have shown that 
the scheme of nonequilibrium thermodynamics also applies to the mesoscopic domain
in which fluctuations strongly influence the dynamics. The
kinetic equations can be obtained from the probability density conservation
law by specifying the phase space diffusion current which follows from the corresponding
entropy production. The theory, called mesoscopic nonequilibrium thermodynamics, has been 
successfully applied to different situations involving
transport \cite{grad temp}, \cite{nosotros} and relaxation phenomena
\cite{chemical}-\cite{review-Rubi}. It is our objective 
in this paper to show its validity
in systems exhibiting memory effects 
by discussing examples pertaining to the domain of nonlinear transport.

The paper is distributed as follows. In Sec. \textbf{2}, we present
the derivation of the Fokker-Planck equation for non-Markovian systems or generalized
Fokker-Planck equation. Section \textbf{3} is devoted to obtain the corresponding
generalized Langevin equation. In Sec. \textbf{4}, we apply previous results
to the case of anomalous diffusion in an external potential. A second application
to the domain of nonlinear transport in ratchet devices is carried
out in Sec. \textbf{5}. Finally, in the discussion Section we summarize
our main results and comment on new perspectives.

\section{The generalized Fokker-Planck equation}

Consider a system whose dynamics is determined by the set of
variables \( \underline{\alpha }\equiv (\alpha _{1},...,\alpha _{n}) \); their
evolution in time \( \underline{\alpha }(t) \) defines a stationary non-Markovian
process. Let \( \underline{\alpha }_{0} \) denote the initial values of the
variables, corresponding to \( t=0 \). The probability of finding
the system in the state \( \underline{\alpha } \) at time \( t \)
is then given by means of the conditional probability density defined
as
\begin{equation}
\label{def. prob. condicional}
P(\underline{\alpha }_{0}|\underline{\alpha },t)\equiv \frac{P(\underline{\alpha }_{0};\underline{\alpha },t)}{P(\underline{\alpha }_{0})},
\end{equation}
 where \( P(\underline{\alpha }_{0};\underline{\alpha },t) \) is the joint
probability density and \( P(\underline{\alpha }_{0}) \) the stationary probability
of the state \( \underline{\alpha }_{0} \), \cite{de groot}. The
probability density satisfies the normalization condition
\begin{equation}
\label{normalizacion}
\int P(\underline{\alpha }_{0}|\underline{\alpha },t)d\underline{\alpha }=1,
\end{equation}
 and is governed by the continuity equation
\begin{equation}
\label{ec. continuidad}
\frac{\partial }{\partial t}P(\underline{\alpha }_{0}|\underline{\alpha },t)=-\frac{\partial }{\partial \underline{\alpha }}\cdot \underline{J}_{\underline{\alpha }}(\underline{\alpha }_{0},\underline{\alpha };t),
\end{equation}
 expressing probability conservation in \( \underline{\alpha } \)-space.
Here \( \underline{J}_{\underline{\alpha }}(\underline{\alpha }_{0},\underline{\alpha };t) \)
represents an unknown probability current.

In order to derive the explicit expression of this current, we will
follow the method proposed by mesoscopic nonequilibrium thermodynamics
(MNET) \cite{beyond},\cite{grad temp}. By assuming local equilibrium
in \( \underline{\alpha } \)-space, we establish the Gibbs equation 
\begin{equation}
\label{ec. gibbs}
\delta S(t)=-\frac{1}{T}\int \mu (\underline{\alpha }_{0},\underline{\alpha };t)\delta P(\underline{\alpha }_{0}|\underline{\alpha },t)d\underline{\alpha },
\end{equation}
 where \( S(t) \) represents the entropy of the system, \( \mu (\underline{\alpha }_{0},\underline{\alpha };t) \)
a nonequilibrium chemical potential and \( T \) the temperature, assumed
constant. For the Markovian case, in which the joint probability factorizes
in the form \( P(\underline{\alpha }_{0};\underline{\alpha },t)=P(\underline{\alpha }_{0})P(\underline{\alpha },t) \),
the Gibbs equation reduces to \cite{grad temp} 
\begin{equation}
\label{gibbs markov}
\delta S(t)=-\frac{1}{T}\int \mu (\underline{\alpha },t)\delta P(\underline{\alpha },t)d\underline{\alpha },
\end{equation}
 where the chemical potential \( \mu (\underline{\alpha },t) \) is related
to the one of the non-Markovian case through 
\begin{equation}
\label{mu markov}
\mu (\underline{\alpha },t)=\int \mu (\underline{\alpha }_{0},\underline{\alpha };t)P(\underline{\alpha }_{0})d\underline{\alpha }_{0},
\end{equation}
 and the probability \( P(\underline{\alpha }_{0}) \) at the initial
state is normalized to the unity.

To determine the value of the current \( \underline{J}_{\underline{\alpha }}(\underline{\alpha }_{0},\underline{\alpha };t) \)
we will first calculate the entropy production \( \sigma (t) \). Taking the
time derivative of Eq. (\ref{ec. gibbs}), substituting Eq. (\ref{ec. continuidad})
and integrating by parts with the condition that the flux vanishes
at the boundaries of the system, we obtain
\begin{equation}
\label{produccion S}
\sigma =-\frac{1}{T}\int \underline{J}_{\underline{\alpha }}(\underline{\alpha }_{0},\underline{\alpha };t)\cdot \frac{\partial }{\partial \underline{\alpha }}\mu (\underline{\alpha }_{0},\underline{\alpha };t)d\underline{\alpha }.
\end{equation}
 Now, following the rules of nonequilibrium thermodynamics we may establish
a linear relationship between the flux and the thermodynamic force in the general
form
\begin{equation}
\label{linear laws}
\underline{J}_{\underline{\alpha }}(\underline{\alpha }_{0},\underline{\alpha };t)=-\frac{1}{T}\int \underline{\underline{L}}(\underline{\alpha },\underline{\alpha }^{\prime };t)\cdot \frac{\partial }{\partial \underline{\alpha }}\mu (\underline{\alpha }_{0},\underline{\alpha }^{\prime };t)d\underline{\alpha }^{\prime },
\end{equation}
 where the Onsager coefficients \( L_{ij}(\underline{\alpha },\underline{\alpha }^{\prime };t) \)
are, in general, function of the state variables \( \underline{\alpha } \) and
satisfy the Onsager reciprocal relations \( \underline{\underline{L}}(\underline{\alpha },\underline{\alpha }^{\prime };t)=\underline{\underline{L^{\dagger }}}(\underline{\alpha }^{\prime },\underline{\alpha };-t) \),
where \( \dagger  \) means the transpose of a matrix \cite{coefonsager}. These
coefficients are related to relaxation functions giving the correlation
of the state variables with their initial conditions \cite{de groot}. In the
Markovian case, correlations decay exponentially and the \( \underline{\underline{L}} \)'s
become independent of time. In general, the \( \underline{\underline{L}} \)'s
may also depend on time through the conditional averages of the state variable
defined by 
\begin{equation}
\label{alfa average}
\underline{\alpha }(t)=\int \underline{\alpha }P(\underline{\alpha }_{0}|\underline{\alpha },t)d\underline{\alpha }.
\end{equation}

To determine the expression of the chemical potential, we will use the
requirement that the Gibbs equation (\ref{ec. gibbs}) has to be compatible
with the Gibbs entropy postulate \cite{de groot}
\begin{equation}
\label{post. gibbs}
S(t)=-k_{B}\int P(\underline{\alpha }_{0}|\underline{\alpha },t)\ln \frac{P(\underline{\alpha }_{0}|\underline{\alpha },t)}{P_{0}(\underline{\alpha })}d\underline{\alpha }+S_{0},
\end{equation}
 where \( k_{B} \) is Boltzmann's constant, \( P_{0}(\underline{\alpha }) \)
is the probability density and \( S_{0} \) the entropy of the reference
state, considered at local equilibrium. By taking variations of this
equation and integrating by parts, we can compare the resulting equation with
Eq. (\ref{ec. gibbs}), obtaining the following expression
for \( \mu (\underline{\alpha }_{0},\underline{\alpha };t) \) 
\begin{equation}
\label{potencial quimico}
\mu (\underline{\alpha }_{0},\underline{\alpha };t)=k_{B}T\ln \frac{P(\underline{\alpha }_{0}|\underline{\alpha },t)}{P_{0}(\underline{\alpha })}+\mu _{0}(\underline{\alpha }).
\end{equation}
 Here \( \mu _{0}(\underline{\alpha }) \) is the chemical potential
at local equilibrium. If we now substitute Eq. (\ref{potencial quimico})
into (\ref{linear laws}), we obtain the constitutive relation for the diffusion
current in \( \underline{\alpha } \)-space
\begin{eqnarray}
\underline{J}_{\underline{\alpha }}(\underline{\alpha }_{0},\underline{\alpha };t)=
-\int \underline{\underline{\tilde{\beta }}}(\underline{\alpha },\underline{\alpha }^{\prime };t)\cdot 
[ k_{B}T\frac{\partial }{\partial \underline{\alpha }^{\prime }}P(\underline{\alpha }_{0}|\underline{\alpha }^{\prime },t). 
\; \; \; \; \; \; \; \; \; \; \; \; \; \; \; \; \; \; \; \; \; \; \; \;  &  & \nonumber \\
+ P(\underline{\alpha }_{0}|\underline{\alpha }^{\prime },t)\underline{X}(\underline{\alpha }^{\prime })] d\underline{\alpha }^{\prime }, & \label{J constitutiva} 
\end{eqnarray}
 where we have introduced the thermodynamic force
\begin{equation}
\label{def. de X}
\underline{X}(\underline{\alpha })\equiv -\frac{\partial }{\partial \underline{\alpha }}[k_{B}T\ln P_{0}(\underline{\alpha })-\mu _{0}(\underline{\alpha })],
\end{equation}
 and defined the matrix  
\begin{equation}
\label{beta non-markovian2}
\underline{\underline{\tilde{\beta }}}(\underline{\alpha },\underline{\alpha }^{\prime };t)\equiv \frac{1}{T}\frac{\underline{\underline{L}}(\underline{\alpha },\underline{\alpha }^{\prime };t)}{P(\underline{\alpha }_{0}|\underline{\alpha },t)}.
\end{equation}

In the Markovian limit, these coefficients must reduce to the friction
coefficients in order to recover the corresponding Fokker-Planck equation from
Eqs. (\ref{ec. continuidad}) and (\ref{J constitutiva}).
As the \( \underline{\underline{L}} \)'s, the \( \underline{\underline{\tilde{\beta }}} \)'s
are invariant under time reversal symmetry \cite{coefonsager} and may also be expressed in terms of 
relaxation functions \cite{de groot}.
Since the entropy production may in general contain contributions from
different irreversible processes cross effects among currents and forces
of identical tensorial nature might arise. Thus, for the sake of simplicity,
the scalar product in Eq. (\ref{J constitutiva}) will implicitly indicate
a sum over all the corresponding contributions and, consequently, \( \underline{\underline{\tilde{\beta }}}(\underline{\alpha },\underline{\alpha }^{\prime };t) \)
is a matrix of matrices containing the coefficients accounting for the possible
cross effects. Now, by substituting Eq. (\ref{J constitutiva}) into (\ref{ec. continuidad})
we finally obtain the generalized Fokker-Planck equation
\begin{eqnarray}
\frac{\partial }{\partial t}P(\underline{\alpha }_{0}| \underline{\alpha },t)=  
\frac{\partial }{\partial \underline{\alpha }}\cdot \int [ k_{B}T\underline{\underline{\tilde{\beta
}}}(\underline{\alpha },\underline{\alpha }^{\prime };t)\cdot \frac{\partial }{\partial \underline{\alpha
}}P(\underline{\alpha }_{0}| \underline{\alpha }^{\prime },t) \; \; \; \; \; \; \; \; \; \; &  & \nonumber \\
 +\underline{\underline{\tilde{\beta }}}(\underline{\alpha },\underline{\alpha }^{\prime };t)\cdot \underline{X}(\underline{\alpha }^{\prime })P(\underline{\alpha }_{0}|\underline{\alpha }^{\prime },t)] d\underline{\alpha }^{\prime }. & \label{FP non-markov} 
\end{eqnarray}
 This equation accounts for the evolution in time of the conditional
probability density for systems exhibiting non-Markovian dynamics when non-local
effects are considered. Eq. (\ref{FP non-markov}) incorporates the memory effects through the time
dependence of the coefficients, \cite{pop-mori},\cite{marcus}. Moreover, it generalizes the non-Markovian Fokker-Planck equation 
obtained in Refs. \cite{adelman}, \cite{cukier} by including nonlinear contributions.

\subsection{The slow variable approximation}

When the characteristic relaxation time scale of the disturbances
produced by thermodynamic forces is short enough when compared with
the corresponding one in the evolution of \( \underline{\alpha } \),
the process becomes local, i.e. \( \underline{\underline{\tilde{\beta }}}(\underline{\alpha },\underline{\alpha }^{\prime };t)= \underline{\underline{\tilde{\beta }}}(\underline{\alpha };t)\delta (\underline{\alpha }^{\prime }-\underline{\alpha }) \).
Under that condition, Eq. (\ref{FP non-markov}) reduces to 
\begin{equation}
\frac{\partial }{\partial t}P(\underline{\alpha }_{0}|\underline{\alpha },t)=\frac{\partial }{\partial
\underline{\alpha }}\cdot \left[ \underline{b}(\underline{\alpha };t)P(\underline{\alpha
}_{0}|\underline{\alpha },t)+
 k_{B}T\frac{\partial }{\partial \underline{\alpha }}\cdot \underline{\underline{\tilde{\beta
}}}(\underline{\alpha };t)P(\underline{\alpha }_{0}|\underline{\alpha },t)\right],  \label{FP Slow}
\end{equation}
 which constitutes the generalized non-Markovian Fokker-Planck equation
in the slow variable approximation \cite{zwanzig libro}. Here we have
defined the general thermodynamic force \( \underline{b}(\underline{\alpha }) \)
by
\begin{equation}
\label{Def. B}
\underline{b}(\underline{\alpha };t)=\underline{\underline{\tilde{\beta }}}(\underline{\alpha };t)\cdot \underline{X}(\underline{\alpha })-k_{B}T\frac{\partial }{\partial \underline{\alpha }}\cdot \underline{\underline{\tilde{\beta }}}(\underline{\alpha };t).
\end{equation}
 The Fokker-Planck equation was previously
derived from the nonlinear Boltzmann equation \cite{kac-logan} and from the 
master equations in \cite{cita a kampi},
respectively. Its Markovian version follows by considering
\( \underline{\underline{\tilde{\beta }}}(\underline{\alpha };t) \) independent
of time. The resulting nonlinear equation agrees with the corresponding
one obtained in \cite{zwanzig libro} from projector operators. In the case
when the force is linear, \( \underline{X}(\underline{\alpha })=\underline{\alpha } \),
and the coefficients only functions of time, Eq. (\ref{FP Slow}) reduces
to
\begin{equation}
\label{FP-markov-lineal}
\frac{\partial }{\partial t}P(\underline{\alpha },t)=
\frac{\partial }{\partial \underline{\alpha }}\cdot \underline{\underline{\tilde{\beta }}}(t)\cdot \left[ \underline{\alpha }P(\underline{\alpha },t)+k_{B}T\frac{\partial }{\partial \underline{\alpha }}P(\underline{\alpha },t)\right] ,
\end{equation}
 where we have used the fact that \( \underline{\underline{\tilde{\beta }}}(t)=\underline{\underline{\tilde{\beta }^{\dagger }}}(t) \).
In particular, for the case of Brownian motion \( \underline{\alpha } \)
is simply the velocity of the particle. When \( \tilde{\beta }_{ij}(\underline{\alpha };t)=\tilde{\beta }(t)\delta _{ij} \),
Eq. (\ref{FP-markov-lineal}) agrees with the generalized Fokker-Planck
equation found in \cite{adelman}.

\subsection{Determination of the coefficients \protect\protect\( \tilde{\beta }_{ij}\protect \protect \)}

The matrix elements \( \tilde{\beta }_{ij} \) can be determined
from the coupled set of constitutive equations (\ref{J constitutiva})
and the evolution equations for the relaxation functions \( \chi _{ij}(\alpha _{i}(t)) \)
defined by
\begin{equation}
\label{Xi}
\chi _{ij}(\alpha _{i}(t))=\int \alpha _{i}\alpha _{0j}\rho (\underline{\alpha }_{0}|\alpha _{i},t)d\alpha _{i},
\end{equation}
 where \( \rho (\underline{\alpha }_{0}|\alpha _{i},t) \)
is the reduced probability
\begin{equation}
\label{densidad n-1}
\rho (\underline{\alpha }_{0}|\alpha _{i},t)\equiv \int P(\underline{\alpha }_{0}|\underline{\alpha },t)d\alpha _{1}...d\alpha _{i-1}d\alpha _{i+1}...d\alpha _{n}.
\end{equation}
 The evolution equations for the \( \chi _{ij} \) are then derived by multiplying
Eq. (\ref{FP Slow}) by \( \alpha _{i} \) and integrating over it. One
obtains the regression laws
\begin{equation}
\label{ec. Xi}
\frac{d}{dt}\chi _{ij}=-\tilde{\beta }_{ij}(t)X_{i}(\underline{\alpha }(t))\alpha _{0j},
\end{equation}
 For arbitrary forces, Eq. (\ref{ec. Xi}) may be approximated by expanding
\( \underline{X}(\underline{\alpha }) \) in \( \underline{\alpha } \). The
coefficients are then given in terms of a cumulant expansion of \( P(\underline{\alpha }_{0}|\alpha _{i},t) \).
In the case of several variables, the remaining conditions for the \( \tilde{\beta }_{ij}(t) \)'s
are obtained by comparison of the different terms entering the hierarchy
of evolution equations for the moments of \( P(\underline{\alpha }_{0}|\underline{\alpha },t) \),
which have been previously approximated, for times \( t\gg \tilde{\beta }^{-1}_{ij}(t) \),
by the corresponding terms obtained from Eqs. (\ref{J constitutiva}) (see
Refs. \cite{nosotros} and \cite{wilemski}). 

The simplest example concerns the case of a Brownian particle in which
\( \alpha _{1}=\vec{u} \). When \( \underline{\underline{\tilde{\beta }}}(t)=\tilde{\beta }(t)\underline{\underline{1}} \),
with \( \underline{\underline{1}} \) the unit matrix, Eq.
(\ref{ec. Xi}) may be written as
\begin{equation}
\label{beta tildl-Xi}
\tilde{\beta }_{\vec{u}\vec{u}}(t)=-\chi _{\vec{u}\vec{u}}^{-1}(t)\frac{d}{dt}\chi _{\vec{u}\vec{u}}(t),
\end{equation}
 where we have defined \( \chi _{\vec{u}\vec{u}}(t) \) by using Eq. (\ref{Xi}).
This formula was used in \cite{adelman} to derive the linear non-Markovian
Fokker-Planck equation. Notice that for Markovian processes \( \chi (\underline{\alpha }(t)) \)
decays exponentially. Then, by using Eq. (\ref{ec. Xi}), one may verify
that \( \tilde{\beta } \) reduces to the friction coefficient \( \beta  \).

\section{The generalized Langevin equation}

Our purpose in this section is to present a new derivation of the generalized
Langevin equation based on our previous formulation. We first start from the
interpretation of the probability density \( P(\underline{\alpha }_{0}|\underline{\alpha },t) \)
as an ensemble average of the density \( \tilde{\rho }(\underline{\alpha }_{0}|\underline{\alpha },t) \)
in the \( \underline{\alpha } \)-space, i.e., \( P(\underline{\alpha }_{0}|\underline{\alpha },t)=\left\langle \tilde{\rho }(\underline{\alpha }_{0}|\underline{\alpha },t)\right\rangle  \).
The actual value of that density differs then from \( P(\underline{\alpha }_{0}|\underline{\alpha },t) \)
in the presence of fluctuations
\begin{equation}
\label{P-fluc}
\delta \tilde{\rho }(\underline{\alpha }_{0}|\underline{\alpha },t)=\tilde{\rho }(\underline{\alpha }_{0}|\underline{\alpha },t)-\left\langle \tilde{\rho }(\underline{\alpha }_{0}|\underline{\alpha },t)\right\rangle .
\end{equation}

Consistently with this interpretation, used in the derivation of the Boltzmann-Langevin
equation \cite{B-L}, Eq. (\ref{ec. continuidad}) results after averaging
the equation

\begin{equation}
\label{continuidad FFP}
\frac{\partial }{\partial t}\tilde{\rho }(\underline{\alpha }_{0}|\underline{\alpha },t)=-\frac{\partial }{\partial \underline{\alpha }}\cdot \underline{\Phi }_{\underline{\alpha }}(\underline{\alpha }_{0},\underline{\alpha };t),
\end{equation}
 with \( \underline{J}_{\underline{\alpha }}(\underline{\alpha }_{0},\underline{\alpha };t)=\left\langle \underline{\Phi }_{\underline{\alpha }}(\underline{\alpha }_{0},\underline{\alpha };t)\right\rangle  \).
The current \( \underline{\Phi }_{\underline{\alpha }}(\underline{\alpha }_{0},\underline{\alpha };t) \)
splits up into systematic and random contributions in the
form
\begin{equation}
\label{Js+Jr}
\underline{\Phi }_{\underline{\alpha }}(\underline{\alpha }_{0},\underline{\alpha };t)=\left\langle \underline{\Phi }_{\underline{\alpha }}(\underline{\alpha }_{0},\underline{\alpha };t)\right\rangle +\underline{\Phi }_{\underline{\alpha }}^{R}(\underline{\alpha }_{0},\underline{\alpha };t).
\end{equation}
 The random part has zero mean and satisfies the fluctuation-dissipation
relation 
\begin{equation}
\label{fluc diss beta} 
\left\langle \underline{\Phi }_{\underline{\alpha }}^{R}(\underline{\alpha }_{0},\underline{\alpha };t)\underline{\Phi }_{\underline{\alpha }}^{R}(\underline{\alpha }_{0},\underline{\alpha }^{\prime };t+\tau )\right\rangle =
2k_{B}T\underline{\underline{\tilde{\beta }}}(\underline{\alpha },\underline{\alpha }^{\prime };\tau )P(\underline{\alpha }_{0}|\underline{\alpha },t)\delta (\tau ). 
\end{equation}
 with the bracket denoting a stationary average and where
we have taken into account Eq. (\ref{beta non-markovian2}). By substituting
(\ref{J constitutiva}) into (\ref{Js+Jr}) and the result into (\ref{continuidad FFP}),
we obtain the fluctuating Fokker-Planck equation
\begin{eqnarray}
\frac{\partial }{\partial t}\tilde{\rho }(\underline{\alpha }_{0}|\underline{\alpha },t)= 
\frac{\partial }{\partial \underline{\alpha }}\cdot \int [ k_{B}T\underline{\underline{\tilde{\beta }}}(\underline{\alpha },\underline{\alpha }^{\prime };t)\cdot \frac{\partial }{\partial \underline{\alpha }}\tilde{\rho }(\underline{\alpha }_{0}|\underline{\alpha }^{\prime },t) +\; \; \; \; \; \; \; \; \; \; \; \; \;  &  & \nonumber \\
+ \underline{\underline{\tilde{\beta }}}(\underline{\alpha }(t))\underline{X}(\underline{\alpha }(t))\tilde{\rho }(\underline{\alpha }_{0}|\underline{\alpha }^{\prime },t)] d\underline{\alpha }^{\prime }-\frac{\partial }{\partial \underline{\alpha }}\cdot \underline{\Phi }_{\underline{\alpha }}^{R}(\underline{\alpha }_{0},\underline{\alpha };t),  \label{FP-fluctuante} 
\end{eqnarray}
 whose average coincides with the generalized Fokker-Planck equation
(\ref{FP non-markov}). In the Markovian case, Eq. (\ref{FP-fluctuante})
reduces to the Fokker-Planck-Langevin equation derived in \cite{mazurMNET}. 

The generalized Langevin equation can now be derived from the
evolution equation of \( \underline{\alpha }(t) \), obtained from
Eq. (\ref{alfa average}), by using in Eq. (\ref{FP-fluctuante}) the
slow variable approximation, and choosing the conditional probability \( \tilde{\rho }(\underline{\alpha }_{0}|\underline{\alpha },t)=\delta (\underline{\alpha }(t)-\underline{\alpha }) \).
After integrating by parts one obtains
\begin{equation}
\label{GLE}
\frac{d}{dt}\underline{\alpha }(t)=-\underline{B}(\underline{\alpha }(t))+\underline{F}^{R}(t),
\end{equation}
 where we have defined the nonlinear force term by 
\begin{equation}
\label{averg force}
\underline{B}(\underline{\alpha }(t))=\underline{\underline{\tilde{\beta }}}(\underline{\alpha }(t))\underline{X}(\underline{\alpha }(t))-k_{B}T\frac{\partial }{\partial \underline{\alpha }(t)}\cdot \underline{\underline{\tilde{\beta }}}(\underline{\alpha }(t)),
\end{equation}
 and the random force by \( \underline{F}^{R}(t)=\int \underline{\Phi }_{\underline{\alpha }}^{R}(\underline{\alpha },t)d\underline{\alpha } \).
Eq. (\ref{GLE}) generalizes the Langevin equation used in Refs. \cite{adelman}
and \cite{cukier} since it incorporates the general force term (\ref{averg force}).
If the process is Markovian, it reproduces the corresponding Langevin
equation since \( \underline{\underline{\tilde{\beta }}}(\underline{\alpha },t)=\underline{\underline{\beta }}(\underline{\alpha }) \)
and, by using \( \tilde{\rho }(\underline{\alpha }_{0}|\underline{\alpha },t)=\delta (\underline{\alpha }(t)-\underline{\alpha }) \),
one obtains \( \underline{\underline{\beta }}(\underline{\alpha }(t))=\int \underline{\underline{\beta }}(\underline{\alpha })\delta (\underline{\alpha }(t)-\underline{\alpha })d\underline{\alpha }, \)
\cite{zwanzig libro}. The fluctuation-dissipation theorem associated to Eq.
(\ref{GLE}) has been derived in the linear non-Markovian
case \cite{pop-mori}:
\begin{equation}
\label{fluc-diss Fr}
\left\langle \underline{F}^{R}(t)\underline{F}^{R}(t+\tau )\right\rangle =-\left( \frac{d}{d\tau }\underline{\underline{\tilde{\beta }}}(\tau )\right) \left\langle \underline{\alpha }(t)\, \underline{\alpha }(t+\tau )\right\rangle ,
\end{equation}
 whereas in the nonlinear Markovian case one has \cite{NLGLE}
\begin{equation}
\label{fluc-diss mazur}
\left\langle \underline{F}^{R}(t)\underline{F}^{R}(t+\tau )\right\rangle =\left\langle \underline{\alpha }\, \underline{B}(\underline{\alpha })+\underline{B}(\underline{\alpha })\, \underline{\alpha }\right\rangle \delta (\tau ),
\end{equation}
 The last equation has been obtained for an arbitrary dependency of the stochastic 
 force on \( \underline{\alpha } \).

The case analyzed in \cite{adelman}, i.e. \( \underline{X}(\underline{\alpha }(t))=\underline{\alpha }(t) \),
follows easily from our formalism. From Eqs. (\ref{GLE}) and (\ref{averg force})
one obtains 
\begin{equation}
\label{GLE-adelman}
\frac{d}{dt}\underline{\alpha }(t)=-\underline{\underline{\tilde{\beta }}}(t)\cdot \underline{\alpha }(t)+\underline{F}^{R}(t),
\end{equation}
 or equivalently to the generalized Langevin equation \cite{zwanzig libro},
\cite{mori} 
\begin{equation}
\label{GLE usual}
\frac{d}{dt}\underline{\alpha }(t)=-\int _{0}^{t}\underline{\underline{\beta }}(t-s)\cdot \underline{\alpha }(s)ds+\underline{F}^{R}(t),
\end{equation}
 where \( \underline{\underline{\beta }}(t) \) is the memory function.
To prove that equivalence, and following \cite{adelman}
we will first take the average of Eqs. (\ref{GLE-adelman}) and (\ref{GLE usual}).
By using Laplace transforms, the solution of Eq. (\ref{GLE usual}) is given
by \( \underline{\alpha }(t)=\underline{\underline{\chi }}(t)\, \underline{\alpha }_{0} \),
where the relaxation function is \( \underline{\underline{\chi }}(t)=L^{-1}\left\{ [s+\underline{\underline{\beta }}(t)]^{-1}\right\}  \).
Taking now the time derivative of \( \underline{\alpha }(t) \),
eliminating \( \underline{\alpha }_{0} \) and using the definition (\ref{beta tildl-Xi}),
we finally arrive at Eq. (\ref{GLE-adelman}).

\section{Brownian motion in an anharmonic heat bath}

In this section, we will analyze the nonlinear dynamics of a Brownian
degree of freedom in an anharmonic bath. 
The state variable will be in this case \( \underline{\alpha }\equiv (\vec{r},\vec{u}) \),
where \( \vec{r} \) is a coordinate and \( \vec{u} \) its conjugate velocity.

Let us consider a reference local equilibrium state characterized by the stationary
probability density 
\begin{equation}
\label{Po B}
P_{0}(\vec{u},\vec{r})=e^{\frac{m}{k_{B}T}\left[ \mu _{0}-\frac{1}{2}u^{2}-\phi \right] },
\end{equation}
where \( \mu _{0} \) is a reference chemical potential per unit mass
and \( \phi (\vec{r}) \) the potential per unit mass imposed on the system  
by the heat bath. 
For this system, the corresponding Gibbs equation is 
\begin{equation}
\label{gibbs thermop}
\delta s=\frac{1}{T}\delta e+\frac{1}{T}p\delta \rho ^{-1}-\frac{1}{T}\int \mu \delta
c_{\vec{u}}d\vec{u}.
\end{equation}
Here \( s \) and \( e \) are the entropy and the internal energy
per unit mass, respectively, \( p \) the hydrostatic pressure, \( c_{\vec{u}}=\frac{mP}{\rho } \)
the Brownian mass fraction and \( \rho \equiv \rho (\vec{u}_{0},\vec{r}_{0}|\vec{r},t) \)
is defined through Eq. (\ref{densidad n-1}). By taking the total time
derivative, \( \frac{D}{Dt}=\frac{\partial }{\partial t}+\vec{v}\cdot \nabla , \)
of Eq. (\ref{gibbs thermop}) and using Eqs. (\ref{Po B}) and (\ref{gibbs thermop}),
we arrive at the entropy production 
\begin{equation}
\label{produccion S brow}
\sigma =-\frac{m}{T}\int \left[ \vec{J}_{\vec{u}}\cdot \frac{\partial \mu }{\partial \vec{u}}+\vec{J}\cdot \nabla (\mu -\phi )+\vec{J}_{\phi }\cdot \nabla \phi \right] d\vec{u}.
\end{equation}
 The average velocity \( \vec{v} \) is defined through the first moment of
the probability density 
\begin{equation}
\label{definicion flujo difusion}
\rho \vec{v}(\vec{r},t)=m\int \vec{u}P(\vec{u}_{0},\vec{r}_{0}|\vec{u},\vec{r},t)d\vec{u}.
\end{equation}

The entropy production \( \sigma  \) adopts the usual form in terms
of flux-force pairs. The two first contributions result from the diffusion current
in \( \vec{u} \)-space \( \vec{J}_{\vec{u}} \) and the current \( \vec{J}\equiv (\vec{u}-\vec{v})P \)
coming up when inertial effects are relevant. The average contributions
arising from the pair \( \vec{J} \) and \( \nabla \phi (\vec{r}) \) vanish
by virtue of Eq. (\ref{definicion flujo difusion}), \cite{de groot}. The third
contribution is genuine of the effect of the potential and introduces the current
\( \vec{J}_{\phi }\equiv \vec{u}P \). In the Markovian case, when external
forces vanish, the second term on the right-hand-side of (\ref{produccion S brow})
contributes to the entropy flux, reproducing in this form the
results of Ref. \cite{inertial}. The constitutive laws giving
the currents \( \vec{J}_{\vec{u}} \), \( \vec{J} \) and
\( \vec{J}_{\phi } \) in terms of the forces are, respectively
\begin{equation}
\label{flujo espacio velocidad}
\vec{J}_{\vec{u}}=-\tilde{\beta }_{\vec{u}\vec{u}}(t)\left\{ \vec{u}P+\frac{k_{B}T}{m}\frac{\partial P}{\partial \vec{u}}\right\} -
\tilde{\beta }_{\vec{u}\vec{r}}(t)\frac{k_{B}T}{m}\nabla P-\tilde{\beta }_{\vec{u}\phi }(t)P\nabla \phi , \label{flujo espacio velocidad} 
\end{equation}
\begin{equation}
\label{flujo u-v} 
\vec{J}=-\tilde{\beta }_{\vec{r}\vec{u}}(t)\left\{ \vec{u}P+\frac{k_{B}T}{m}\frac{\partial P}{\partial \vec{u}}\right\} -
\tilde{\beta }_{\vec{r}\vec{r}}(t)\frac{k_{B}T}{m}\nabla P-\tilde{\beta }_{\vec{r}\phi }(t)P\nabla \phi , 
\end{equation}
 and 
\begin{equation}
\label{flujo v}
\vec{J}_{\phi }=-\tilde{\beta }_{\phi \vec{u}}(t)\left\{ \vec{u}P+\frac{k_{B}T}{m}\frac{\partial P}{\partial \vec{u}}\right\} -
\tilde{\beta }_{\phi \vec{r}}(t)\frac{k_{B}T}{m}\nabla P-\tilde{\beta }_{\phi \phi }(t)P\nabla \phi ,  
\end{equation}
 where the Onsager coefficients have been defined consistently with
Eq. (\ref{beta non-markovian2}). By substituting Eq. (\ref{flujo espacio velocidad})
into the corresponding continuity equation, one obtains the generalized
Fokker-Planck equation describing the non-Markovian dynamics of the
Brownian degree of freedom in the presence of arbitrary bounding forces
\begin{eqnarray}
\frac{\partial P}{\partial t}+\nabla \cdot \vec{u}P=  
\frac{\partial }{\partial \vec{u}}\cdot \left\{ \tilde{\beta }_{\vec{u}\vec{u}}(t)\left[
\vec{u}P+\frac{k_{B}T}{m}\frac{\partial P}{\partial \vec{u}}\right] \right. -\; \; \; \; \; \; \; \;\; \;  &  & \nonumber \\
-\left. \tilde{\beta }_{\vec{u}\phi }(t)\vec{X}(\vec{r})P+\frac{k_{B}T}{m}\tilde{\beta }_{\vec{u}\vec{r}}(t)\nabla P\right\} , &  & \label{F-P para TNL} 
\end{eqnarray} 
where \( \vec{X}(\vec{r})=-\nabla \phi  \). In the linear
case, it reduces to the corresponding one obtained through the conditional
probability associated to the Langevin equation used in \cite{adelman}. As
a consequence of the existence of local equilibrium in \( \vec{u} \)-space, \( \tilde{\beta }_{\vec{u}\vec{u}}(t) \)
multiplies both the drift and the diffusion terms entering (\ref{F-P para TNL}). 
The presence of forces adds two terms that make the 
fluctuation-dissipation theorem not longer valid. 
For Markovian processes in which \( \tilde{\beta }_{\vec{u}\vec{u}}(t)=\beta  \)
, \( \tilde{\beta }_{\vec{u}\phi }(t)=1 \) and \( \tilde{\beta }_{\vec{u}\vec{r}}(t)=0 \),
Eq. (\ref{F-P para TNL}) reproduces the corresponding Fokker-Planck
equation \cite{zwanzig libro}. By integrating Eqs. (\ref{flujo u-v}) and
(\ref{flujo v}) over \( \vec{u} \) and combining the resulting equations,
we obtain the diffusion current \( \vec{J}_{D}(\vec{r},t)\equiv \rho \vec{v}(\vec{r},t) \) 
\begin{equation}
\label{JD activated}
\vec{J}_{D}=-\tilde{\eta }(t)\left[ \frac{k_{B}T}{m}\nabla \rho -\rho \vec{X}(\vec{r})\right] ,
\end{equation}
where the existence of local equilibrium imposes \( \tilde{\eta }(t)\equiv \tilde{\beta }_{\vec{u}\phi }^{-1}[ \tilde{\beta }_{\phi \phi }(t)-\tilde{\beta }_{\vec{r}\phi }(t)\tilde{\beta }_{\vec{r}\vec{u}}^{-1}(t)] =\tilde{\beta }_{\vec{u}\phi }^{-1}[ \tilde{\beta }_{\phi \vec{r}}(t)-\tilde{\beta }_{\vec{r}\vec{r}}(t)\tilde{\beta }_{\vec{r}\vec{u}}^{-1}(t)]  \).
By averaging Eq. (\ref{F-P para TNL}) over \( \vec{u} \)-space and substituting
Eq. (\ref{JD activated}) into the resulting equation, we obtain the non-Markovian
Smoluchowski equation for the density field \( \rho  \) 
\begin{equation}
\label{Smoluch Oxtoby}
\frac{\partial \rho }{\partial t}=D(t)\nabla ^{2}\rho -\tilde{\eta }(t)\nabla \cdot \left( \rho \vec{X}(\vec{r})\right) ,
\end{equation}
where we have defined the diffusion coefficient \( D(t) \) in terms of the
generalized Einstein relation 
\begin{equation}
\label{coef. dif. smoluch}
D(t)=\frac{k_{B}T}{m}\tilde{\eta }(t),
\end{equation}
from which it follows that the slow modes of non-Markovian systems
satisfy a modified fluctuation-dissipation relation. For the particular
case of a harmonic potential, \( \phi (\vec{r})=\frac{1}{2}\omega ^{2}r^{2} \),
where \( \omega  \) is the characteristic frequency, Eqs. (\ref{F-P para TNL})
and (\ref{Smoluch Oxtoby}) reproduce the generalized Fokker-Planck and Smoluchowski
equations derived in \cite{adelman} and \cite{oxtoby 1}, respectively. Now,
multiplying Eq. (\ref{F-P para TNL}) by \( \vec{u} \) and integrating over
it, we obtain the evolution equation of the average velocity \( \vec{v} \)
\begin{equation}
\label{evolucion momentum}
\rho \frac{D\vec{v}}{Dt}=-\tilde{\beta }_{\vec{u}\vec{u}}(t)\left[ \rho \vec{v}-\tilde{\eta }(t)\rho \vec{X}(\vec{r})+D(t)\nabla \rho \right] ,
\end{equation}
 where we have assumed that the pressure tensor of the ideal Brownian
gas is given by \( \vec{\vec{P}}=\frac{k_{B}T}{m}\rho \vec{\vec{1}} \). For
times satisfying \( \tau \gg \tilde{\beta }_{ij} \), we have imposed that Eq.
(\ref{evolucion momentum}) must coincide with Eq. (\ref{JD activated}), obtaining
the conditions \( \tilde{\beta }_{\vec{u}\phi }(t)=\tilde{\beta }_{\vec{u}\vec{u}}(t)\tilde{\eta }(t) \)
and \( \tilde{\beta }_{\vec{u}\vec{r}}(t)=\tilde{\beta }_{\vec{u}\phi }(t)-1 \),
consistent with the corresponding ones for the Markovian case. \( \tilde{\beta }_{\vec{u}\vec{u}}(t) \)
and \( \tilde{\eta }(t) \) may be obtained by using the regression
law (\ref{ec. Xi}) and Eqs. (\ref{F-P para TNL}) and (\ref{Smoluch Oxtoby}).

When local equilibrium in $\vec{r}$-space is not fulfilled, the coefficients appearing in the 
diffusion and drift terms of Eqs. (\ref{flujo u-v}), (\ref{flujo v}) and (\ref{JD activated}) 
must be found independently. Then, even in the absence of external forces, the 
fluctuation-dissipation theorem (\ref{coef. dif. smoluch}) is no longer valid. This fact
occurs in systems exhibiting anomalous diffusion, which are described by non-Markovian 
Fokker-Planck and Langevin equations, \cite{tokuyama} and \cite{brasil}.

\section{Ratchet effect induced by Onsager couplings}

In this section, we will apply our formalism to show a new manifestation
of the ratchet effect which originates from the Onsager coupling of the probability
current and a periodic thermodynamic force, when fluctuations are rectified
by an external asymmetric potential. We will show that in the scale considered 
the non-Markovian dynamics makes the ratchet effect more pronounced. 

The continuity equation in the presence of external forces takes the form 
\begin{equation}
\label{ec. continuidad ratchet}
\frac{\partial P}{\partial t} + \nabla \cdot \vec{u}P + \vec{F} \cdot \frac{\partial P}{\partial \vec{u}}=-\frac{\partial }{\partial \vec{u}}\cdot \vec{J}_{\vec{u}},
\end{equation}
where the external force per unit mass is $\vec{F}(\vec{r}) = - \nabla V(\vec{r})$, where $V(\vec{r})$ 
represents the ratchet potential. In the present case, we will consider that the probability 
density of the reference 
state is given by Eq. (\ref{Po B}) with $\phi (\vec{r})=0$. 

Inhomogeneities in the temperature of 
the bath induce a heat flux \( \vec{J}_{q} \) whose appearance modifies the entropy 
production of the system. In this case, the Gibbs entropy postulate (\ref{post. gibbs}) and 
the Gibbs equation (\ref{gibbs thermop}) must be complemented with the balance 
equation for the internal energy \( e \)
\begin{equation}
\label{balance energia thermop}
\frac{De}{Dt}=-\nabla \cdot \vec{J}_{q},
\end{equation}
 where viscous effects have been neglected. By considering the Gibbs-Duhem
relation \cite{de groot}, the entropy production can be written in the form
\begin{equation}
\label{prod. S thermop} 
\sigma =-\vec{J}^{*}_{q}\cdot \frac{1}{T^{2}}\nabla T-\frac{m}{T}\int \vec{J}_{\vec{u}}\cdot \frac{\partial \mu }{\partial \vec{u}} d\vec{u},  
\end{equation}
 where the modified heat flow \( \vec{J}_{q}^{*} \) is given by
\begin{equation}
\label{Jq modified}
\vec{J}_{q}^{*}=\vec{J}_{q}-\frac{1}{2}m\int u^{2}\vec{J}\, d\vec{u}.
\end{equation}
 The second term on the right-hand-side of Eq. (\ref{Jq modified}) represents
a local flow of kinetic energy. 

The diffusion current \( \vec{J}_{\vec{u}} \)
is, according with Onsager theory coupled to the temperature gradient and is
given by 
\begin{equation}
\label{Ju thermop} 
\vec{J}_{\vec{u}}=-\tilde{\beta }_{\vec{u}\vec{u}}(t)\left\{ \vec{u}P+\frac{k_{B}T}{m}\frac{\partial P}{\partial \vec{u}}\right\} -\tilde{\beta }_{\vec{u}T}(t)P\frac{\nabla T}{T}.  
\end{equation}
 Similarly, the heat current can be formulated as 
\begin{equation}
\label{Jq thermop} 
\vec{J}^{*}_{q}=-\tilde{\beta }_{TT}(t)\nabla T- \tilde{\beta }_{T\vec{u}}(t)\rho \vec{v},
\end{equation}
 where  \( \tilde{\beta }_{TT}(t)\equiv \frac{L_{TT}(t)}{T^{2}} \) ,
and the coupling coefficient  \( \tilde{\beta }_{T\vec{u}} \) 
has been defined accordingly with
Eq. (\ref{Def. B}). By substituting Eq. (\ref{Ju thermop}) into the corresponding
continuity equation for the probability density, we obtain the generalized Fokker-Planck
equation
\begin{eqnarray}
\frac{\partial P}{\partial t}+\nabla \cdot \vec{u}P + \vec{F} \cdot \frac{\partial P}{\partial
\vec{u}} \; \; \; \; \; \; \; \; \; \; \; \; \; \; \; \; \; \; \; \; \; \; \; \; \; \; \; \; \; \; \; \; \; \; \; \; \; \; \; \;  &  & \nonumber \\
=\frac{\partial }{\partial \vec{u}}\cdot \left\{ \tilde{\beta }_{\vec{u}\vec{u}}(t)\left[
\vec{u}P+\frac{k_{B}T}{m}\frac{\partial P}{\partial \vec{u}}\right] + \tilde{\beta
}_{\vec{u}T}(t)P\frac{\nabla T}{T}\right\}, & & \label{F-P thermop} 
\end{eqnarray}
which describes the transport of a Brownian particle in an thermally inhomogeneous environment.

By multiplying Eq. (\ref{F-P thermop}) by \( \vec{u} \) and integrating over
it, we obtain the corresponding evolution equation for the momentum
\begin{equation}
\label{Jtd}
\rho \frac{D\vec{v}}{Dt}=-\tilde{\beta }_{\vec{u}\vec{u}}(t)\left[ \rho \vec{v}+D(t)\nabla \rho \right. 
\left. +D_{T}(t)\rho \nabla T\right]+\rho \vec{F}(\vec{r}) ,  
\end{equation}
 where \( \tilde{\beta }_{\vec{u}\vec{u}}(t) \) has been given in Eq. (\ref{beta tildl-Xi}) 
and we have defined the diffusion coefficient \( D(t)= \frac{k_{B}T}{m}\tilde{\beta }^{-1}_{\vec{u}\vec{u}}(t) \)
and the thermal diffusion coefficient
\( D_{T}(t)=\frac{D(t)}{T}\left( 1+\frac{m}{k_{B}T}\tilde{\beta }_{\vec{u}T}(t)\right)  \). For 
convenience, we will assume
that \( D(t) \) and \( D_{T}(t) \) will be only functions of time. The diffusion
current \( \vec{J}_{D} \) is obtained from Eq. (\ref{Jtd}) at times \( t\gg \tilde{\beta }^{-1}_{\vec{u}\vec{u}}(t) \).
By following the steps indicated in Sec. \textbf{4}, we arrive at
the generalized Smoluchowski equation
\begin{equation}
\label{Smoluch thermop}
\frac{\partial \rho }{\partial t}=D(t)\nabla ^{2}\rho +D_{T}(t)\nabla \cdot \left( \rho \nabla
T\right) -\tilde{\beta }^{-1}_{\vec{u}\vec{u}}(t)\nabla \cdot \left[ \rho \vec{F}(\vec{r})\right] ,
\end{equation}
 where memory effects enter the equation through the time dependence of \( D(t) \),
\( D_{T}(t) \) and \( \tilde{\beta }_{\vec{u}\vec{u}}(t) \). Since in practice the temperature
inhomogeneities originated by density gradients are negligible \cite{de groot},
by substituting the diffusion current, implicit in Eq. (\ref{Smoluch thermop}),
and Eqs. (\ref{Jq modified}) and (\ref{Jq thermop}) into (\ref{balance energia thermop}),
we arrive at the heat conduction equation
\begin{equation}
\label{Ec. T}
c_{v}\frac{\partial T}{\partial t}=\tilde{\lambda }_{eff}(t)\nabla ^{2}T,
\end{equation}
 where $c_{v}$ is the heat capacity. We have used the relation, \( e=c_{v}T \) and defined the
effective heat conduction coefficient \( \tilde{\lambda }_{eff}(t)=\tilde{\beta
}_{TT}(t)\left[1+ \rho D_{T}(t)\frac{\tilde{\beta }_{\vec{u}T}(t)}{\tilde{\beta }_{TT}(t)}\right]  \).
To obtain Eq. (\ref{Ec. T}), we have also neglected contributions arising
from the kinetic energy flow \cite{nosotros}, and assumed temperature gradients
such that \( \left( \frac{l_{0}\nabla T}{T}\right) ^{2}\ll 1 \), where
\( l_{0} \) is a characteristic length of the bath. For $\tilde{\beta }_{TT}(0)$ large 
enough ($\sim 10^{4}gr\;cm\;s^{-3}K^{-1}$) and 
$D_{T}(0), \tilde{\beta }_{\vec{u}T}(0)\ll 1$, $\tilde{\lambda }_{eff}$ is practically constant
in first approximation.

To illustrate the ratchet effect, we consider the case of a Brownian particle moving in 
a bath, filling the space $x>0$, in
which a periodic temperature gradient is established when \( T(0,t)=T_{0}[ 1+\cos (\omega t)] \). This situation may be accomplished
when the distance of the particle to the wall is smaller than the wave length of the
thermal waves. In such a case the spatial dependence in the propagating wave 
\begin{equation}
\label{T(x,t)}
T(x,t)=T_{0}\left[ 1+e^{-kx} \cos (kx-\omega t)\right],
\end{equation}
 solution of Eq. (\ref{Ec. T}), can be neglected. Here \( \omega  \) is a frequency and \(
k=(\frac{\omega c_{v}}{2 \lambda_{eff}})^{\frac{1}{2}} \) a wave number. The temperature gradient is then given by
$\nabla T=k T_{0}[\sin(\omega t) - \cos (\omega t)]$. This approximation is valid for
values of the parameters indicated in the figure captions, all of them accessible in 
experiments.
\begin{figure}[H]
\begin{center}
\includegraphics[scale=.49]{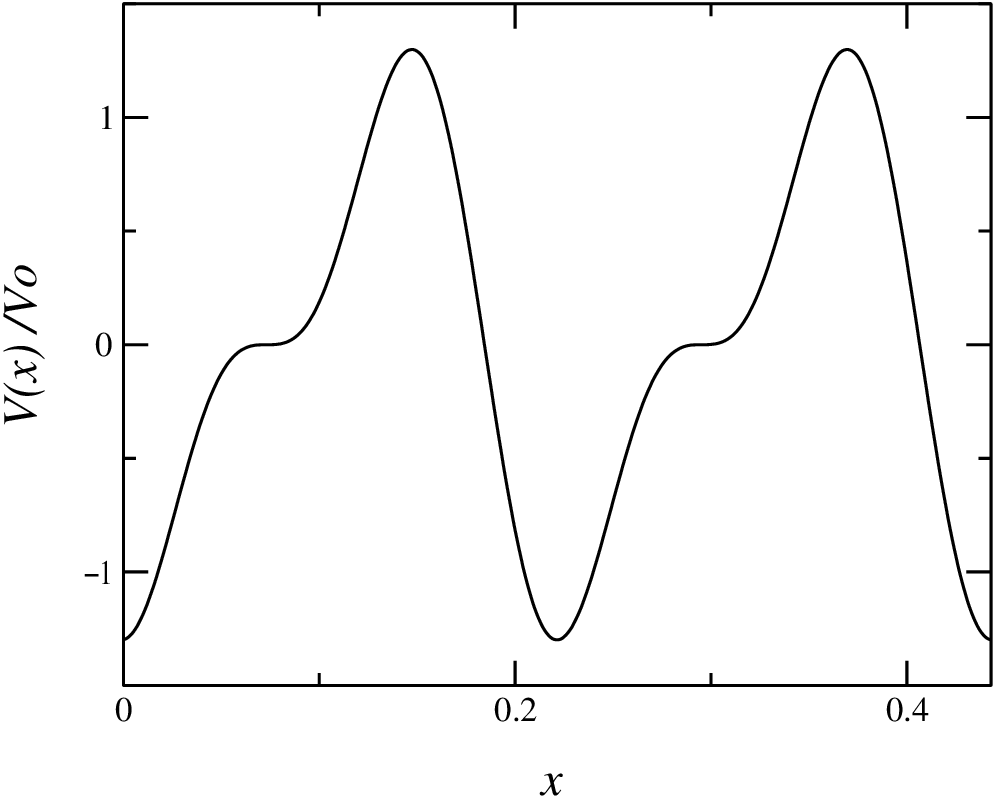}
\caption{{\small Ratchet potential: $V(x)=-\frac{V_{0}}{2\pi }[ \sin (9\pi x)
+\frac{1}{2}\sin (18\pi x)]$; $V_{0}$ is a constant.}}
\end{center}
\end{figure}
 The particle in turn is under the influence of the asymmetric
potential (see Fig. 1). We will assume that it undergoes anomaluos diffusion with
\( D(t)\sim t^{\frac{1}{2}} \) and \( \tilde{\beta }_{\vec{u}\vec{u}}(t)\sim t^{-\frac{1}{2}} \).
This particular dependence has recently been observed for the Brownian motion
of a particle inside an eukaryotic cell \cite{cellular transport}. The diffusion
coefficient \( D(t) \) has been obtained through the time derivative of the mean square displacement measured
in \cite{cellular transport}, and \( \tilde{\beta }_{\vec{u}\vec{u}}(t) \)
by using Eq. (\ref{ec. Xi}). The thermal diffusion coefficient
\( D_{T}(t) \)
is then determined by assuming that the Soret coefficient 
\( S_{T}\equiv \frac{D_{T}(t)}{D(t)} \) is constant in first approximation.
According with experimental results \cite{soret-exps}, we will consider that \( D(0)\sim 10^{-7}cm^{2}s^{-1} \)
and that \( S_{T}\sim 10^{-1}K^{-1} \). Then, by using \( \rho (x,t)=\delta (x(t)-x) \)
in Eq. (\ref{Jtd}), after integrating over the system volume we obtain the
evolution equation for the position of the particle
\begin{equation}
\label{x-punto(t)}
\frac{d^{2}x(t)}{dt^{2}}+\tilde{\beta }_{\vec{u}\vec{u}}(t)\frac{dx(t)}{dt}=F(x(t))- 
\tilde{\beta }_{\vec{u}\vec{u}}(t)D_{T}(t)\left[ \frac{\partial T}{\partial x}\right]
_{x=x(t)}+F^{R}(t). 
\end{equation}
 The inertia term is maintained since non-Markovian effects arise at short time scales. 

Positions and velocities of the Brownian particle have been determined from
this equation by using a fourth order Runge-Kutta method.
In Fig. 2, we show the displacement as a function of time for a given initial
condition and a time periodic force. After a transient regime determined by 
$\tilde{\beta }_{\vec{u}\vec{u}}(t)$, in the non-Markovian case 
the current, proportional to the slope of the curve, becomes preactically constant. The average 
motion has the same frequency as the time periodic force. 
\begin{figure}[H]
\begin{center}
\includegraphics[scale=.49]{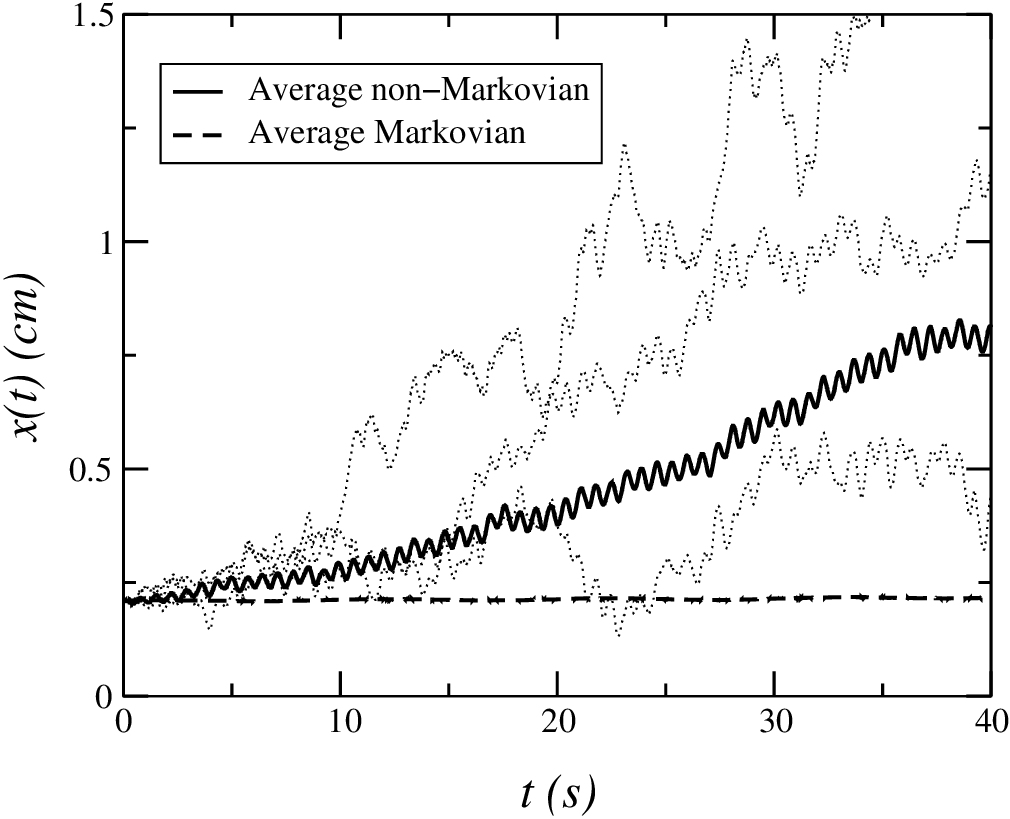}
\caption{{\small Position of the Brownian particle versus time. Dotted lines represent  
different realizations of the noise, whereas the solid line corresponds to the average 
displacement of the Brownian particle for 20 realizations of the noise in the 
non-Markovian case. The dashed line represents the average position in the Markovian 
case. The values of the different parameters are:
$F(0)=1\times10^{-3}gr\;cm^{2}s^{-2}$, $\nabla T\simeq1.3 Kcm^{-1}$, 
\( \omega =9\times10^{-3}s^{-1} \), $k\simeq2.1\times10^{-2}cm^{-1}$ and 
$m\simeq 10^{-4}gr$.}}
\end{center}
\end{figure}

In Fig. 3, we represent average displacements
for different values of the parameters. In curves a) and b) the strength of the potential and the 
frequency of temperature gradient have been used to control the magnitude of 
the current. We have 
verified that no net current appears
when the potential is symmetric. As a consequence of the fact that the damping 
coefficient decreases with time, we may conclude that, under similar conditions,
the non-Markovian system is more efficacious in rectifying the motion than the 
Markovian one.

\begin{figure}[H]
\begin{center}
\includegraphics[scale=.49]{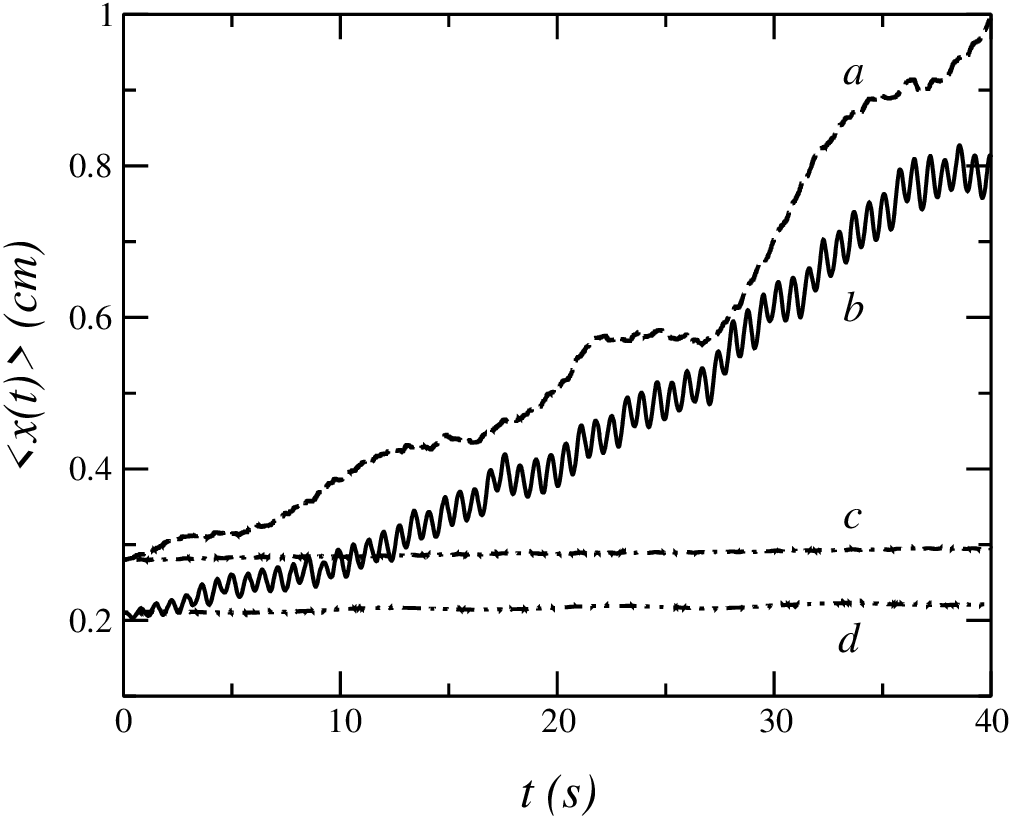}
\caption{{\small Average displacements for different initial positions, frequencies and strengths of the temperature 
gradient with $F(0)=6.8\times10^{-3}gr\;cm^{2}s^{-2}$, and $m\simeq 10^{-4}gr$. In cases 
 $a)$ and $c)$: $\nabla T\simeq 1.3Kcm^{-1}$ and \( \omega =9\times
10^{-3}s^{-1} and\; k\simeq2.1\times
10^{-3}cm^{-1}\). In cases  
$b)$ and $d)$: $\nabla T\simeq 0.2Kcm^{-1}$, \( \omega =6\times 10^{-4}s^{-1} and\; 
 k\simeq .1\times 10^{-4}cm^{-1}\). Curves $a$ and $b$ correspond to non-Markovian cases,
  whereas $c$ and $d$ result from Markovian dynamics.
}}
\end{center}
\end{figure}

\section{Discussion}

In this paper, we have proposed a new method to derive the generalized Fokker-Planck
equation describing the dynamics of systems when memory effects are relevant.
Our formalism is able to derive kinetic equations from the probability density conservation
law by providing expressions for the probability current obtained from the entropy
production in phase space. It permits to incorporate in a simple way basic ingredients
proper of systems outside equilibrium as the presence of external forces and
of inhomogeneities in the bath.

We have presented two applications. In the first one, we have discussed the 
non-Markovian Brownian motion in a bath exerting a force on the 
Brownian particle
\cite{oxtoby 1}, \cite{adelman}. The Fokker-Planck and Smoluchowski equations
have been derived for an unspecified force by avoiding the difficulties inherent
to the habitual method, obtaining them from the corresponding Langevin equation.
Our method relates the validity of the fluctuation-dissipation theorem
to the existence of local equilibrium in the space of the mesoscopic variables.
An example in which this feature becomes manifest is the case of anomalous
diffusion \cite{tokuyama},\cite{brasil}. 
In the second example, we have analyzed the role played by the
presence of inhomogeneities in the bath, due to an imposed temperature gradient,
in nonlinear transport. We have found
a new ratchet effect which originates from the cooperation of a 
thermodynamic periodic force with the nonlinear potential. The positiveness of the entropy
production imposes restrictions on the values of the parameters entering mathematical
models.

These results confirm the usefulness of our method which offers a very promising
framework in the study of systems operating at the mesoscale for which the presence
of nonlinearities, the participation of many degrees of freedom and the influence
of inhomogeneities in the environment become a common feature of the dynamics.

\section{ACKNOWLEDGMENTS}

We want to acknowledge Victor Romero-Roch\'{\i}n, D. Reguera, A. P\'{e}rez-Madrid, J. M.
Vilar, I. Sokolov and M. Bogunya for fruitful discussions. I.S.H. acknowledges UNAM-DGAPA for economic 
support. This work was partially supported by DGICYT of the Spanish Government under 
Grant No. PB2002-01267.

\end{document}